\documentclass[]{svmult}
\usepackage{type1cm}         
\usepackage{graphicx}        
\usepackage[bottom]{footmisc}

\usepackage{newtxtext}       
\usepackage[varvw]{newtxmath}

\usepackage{marvosym,amsmath,empheq}

\DeclareMathOperator{\Shi}{Shi}
\DeclareMathOperator{\Si}{Si}
\newcommand*{\defeq}{\stackrel{\text{def}}{=}}

\usepackage[square,numbers]{natbib}

\usepackage[colorlinks,citecolor=blue]{hyperref}
\usepackage{bookmark}
\makeatletter
\def\toclevel@title{-1}
\def\toclevel@author{0}
\makeatother

\begin{document}

\title*{Series solutions for clamped peridynamic beams using fourth-order eigenfunctions}
\titlerunning{Series solutions for peridynamic beams}

\author{Ziyu Wang\protect\texorpdfstring{\orcidID{0000-0002-7749-3152}}{} and\protect\texorpdfstring{\\}{} Ivan C.\ Christov\protect\texorpdfstring{\orcidID{0000-0001-8531-0531}}{}}

\authorrunning{Z.\ Wang and I.\ C.\ Christov}

\institute{Ziyu Wang \at School of Mechanical Engineering, Purdue University, West Lafayette, Indiana 47907, USA \email{wang5600@purdue.edu}
\and Ivan C.\ Christov (\Letter) \at School of Mechanical Engineering, Purdue University, West Lafayette, Indiana 47907, USA \email{christov@purdue.edu}}

\maketitle

\abstract{We propose an analytical approach to solving nonlocal generalizations of the Euler--Bernoulli beam. Specifically, we consider a version of the governing equation recently derived under the theory of peridynamics. We focus on the clamped--clamped case, employing the natural eigenfunctions of the fourth derivative subject to these boundary conditions. Static solutions under different loading conditions are obtained as series in these eigenfunctions. To demonstrate the utility of our proposed approach, we contrast the series solution in terms of fourth-order eigenfunctions to the previously obtained Fourier sine series solution. Our findings reveal that the series in fourth-order eigenfunctions achieve a given error tolerance (with respect to a reference solution) with ten times fewer terms than the sine series. The high level of accuracy of the fourth-order eigenfunction expansion is due to the fact that its expansion coefficients decay rapidly with the number of terms of the series, one order faster than the Fourier series in our examples.}


\section{Introduction}
\label{sec:intro}

Peridynamics is a recent reformulation of continuum mechanics introduced by Silling \cite{silling2000,silling2010} to overcome the limited applicability of classical continuum mechanics to problems with discontinuities. The idea is that any material point in the continuum can interact with other points within a finite \emph{horizon} of itself, giving peridynamic theory a ``nonlocal'' character. Different approaches to nonlocality are also discussed in the context of generalized continua, see, e.g., \cite{altenbach2011}. Due to peridynamic theory's ability to model the initiation and growth of cracks in materials without the need for additional techniques to handle such discontinuities, peridynamics is often used to simulate the deformation of elastic materials until failure. For example, \citet{kim2023} simulated the deformation, damage, and failure of an elastic band under fluid flow loading in a fluid--structure interaction framework. They demonstrated that a peridynamics-enabled computational model could capture how the flow forces not only deform but also tear apart the elastic band, a simulation that would not be possible within a fluid--structure interaction framework based on classical continuum mechanics. Although peridynamics theory is typically implemented as a computational framework, there is nevertheless fundamental interest in finding analytical solutions of the governing equations of peridynamics and peristatics \cite{mikata2012,nishawala2017,huang2018,mikata2023}.

Deriving reduced models, such as for beams, plates, and shells, is a time-honored research area within classical continuum mechanics \cite{timoshenko1959,kraus1967,altenbach2004}. More recently, there has been an interest in doing the same for peridynamic beams \cite{ogrady2014a,yang2020,yang2022a,yang2023a,yang2023b}, membranes \cite{silling2005}, plates \cite{ogrady2014b,diyaroglu2015,taylor2015,yang2022b,naumenko2022}, and shells \cite{ogrady2014b}. Once such models are developed, it is of interest to find analytical solutions to basic problems (if possible) \cite{challamel2018,yang2020,yang2022a,yang2022b,yang2023a,yang2023b}. In this work, we continue the quest, providing new analytical peristatic solutions for the bending of clamped--clamped beams using fourth-order eigenfunctions.

Perhaps for obvious reasons, the eigenfunctions of the fourth-derivative operator are often termed \emph{beam} functions (e.g., \citep{papanicolaou2003,papanicolaou2009}). Historically, Lord Rayleigh \citep[\S\S170-175]{strutt1877} derived these functions in the context of the lateral vibrations of elastic bars, including clamped--clamped, clamped--free, and loaded boundary conditions amongst other combinations. \citet[\S133]{chandrasekhar1981} also derived these orthogonal eigenfunctions and employed them to study the linear stability of thermal convection.

Motivated by the recent work of \citet{yang2022a,yang2022b,yang2023a}, we continue the search for closed-form analytical solutions to peridynamic problems. \citet{yang2023a} used Castigliano's theorem to replace the clamped--clamped boundary conditions with moment conditions on a supported beam, a case that can be solved by a Fourier sine series. On the other hand, in this work, we demonstrate that the fourth-order (beam) eigenfunctions, which naturally satisfy the clamped boundary conditions, can also be used to obtain analytical series solutions without resorting to Castigliano's theorem. Moreover, we show that the series in terms of fourth-order eigenfunctions converges significantly faster than a Fourier sine series, which means the series solutions derived by our approach provide a more accurate representation of peridynamic beam bending (for a fixed number of terms in the series).


\section{The fourth-order ``beam'' eigenfunctions}
\label{sec:beam_efuncs}

\begin{figure}[t]
    \centering
    \includegraphics[width=.9\textwidth]{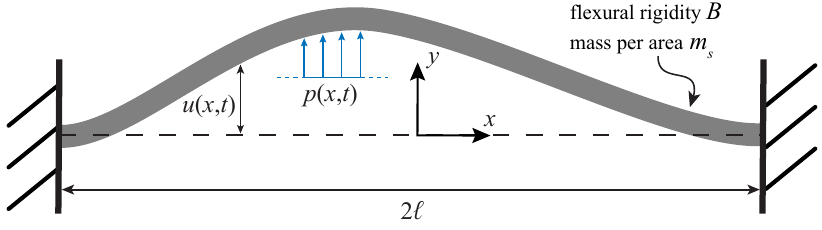}
    \caption{Schematic of a clamped--clamped beam bending due to a pressure load.}
    \label{fig:beam}
\end{figure}

First, we review the problem of the dynamic bending of an Euler--Bernoulli beam, originating from ``classical continuum mechanics,'' as depicted in the schematic in Fig.~\ref{fig:beam}. The beam has mass per area $m_s$, bending rigidity $B$, and applied pressure load $p(x,t)$. When it is clamped at both ends, its vertical displacement $u(x,t)$ obeys the fourth-order initial-boundary-value problem (IBVP):
\begin{subequations}\label{eq:dim_beam_ibvp}\begin{empheq}[left = \empheqlbrace]{alignat=3}
    m_s \frac{\partial^2 u}{\partial t^2} + B \frac{\partial^4 u}{\partial x^4} &= p(x,t), &\qquad -\ell<x<\ell,\quad t>0,\label{eq:dim_beam_eq}\\[1.2mm]
    u|_{x=\pm\ell} = \left.\frac{\partial u}{\partial x}\right|_{x=\pm \ell} &= 0, &\qquad t>0,\\
    u(x,t) &= u^0(x), &\qquad -\ell \le x \le \ell.
\end{empheq}\end{subequations}
We make the IBVP~\eqref{eq:dim_beam_ibvp} dimensionless by introducing the following variables based on the typical load scale $p_0$:
\begin{multline}
    T = \frac{t}{\sqrt{m_s\ell^4/B}},\quad X = \frac{x}{\ell},\quad 
    U(X,T) = \frac{u(x,t)}{p_0\ell^4/B}, \quad P(X,T) = \frac{p(x,t)}{p_0},
\end{multline}
to obtain
\begin{subequations}\label{eq:beam_ibvp}\begin{empheq}[left = \empheqlbrace]{alignat=3}
    \frac{\partial^2 U}{\partial T^2} + \frac{\partial^4 U}{\partial X^4} &= P(X,T), &\qquad -1 < X < 1,\quad t>0, \label{eq:nd_beam_eq}\\[1.2mm]
    U|_{X=\pm1} = \left.\frac{\partial U}{\partial X}\right|_{X=\pm1} &= 0, &\qquad T>0, \label{eq:clamped_bc}\\
    U(X,T) &= U^0(X), &\qquad -1 \le X \le 1. \label{eq:ic}
\end{empheq}\end{subequations}
IBVP~\eqref{eq:beam_ibvp} is referred to as ``classical beam theory'' (CBT) in previous works.

Following the notation of Papanicolaou et al.~\citep{papanicolaou2003,papanicolaou2009}, the fourth-order Sturm--Liouville eigenvalue problem (EVP) on $X\in[-1,+1]$ associated to the IBVP~\eqref{eq:beam_ibvp} is
\begin{subequations}\label{eq:evp}\begin{empheq}[left = \empheqlbrace]{alignat=3}
    +\frac{d^4 \psi}{d X^4} &= \lambda^4 \psi, &\qquad -1 < X < 1,\\[1.2mm]
    \psi|_{X=\pm1} &= \left.\frac{d \psi}{d X}\right|_{X=\pm1} = 0.	   
\end{empheq}\end{subequations}
The EVP~\eqref{eq:evp} is self-adjoint and there exist two linearly independent sets of eigenfunctions $\{\psi^s_m,\psi^c_m\}$ with associated eigenvalues $\{\lambda^s_m,\lambda^c_m\}$ \cite{papanicolaou2003,papanicolaou2009}:
\begin{subequations}\label{eq:efuncs}\begin{empheq}[left = \empheqlbrace]{alignat=3}
    \psi^s_m(X) &= \frac{1}{\sqrt{2}}\left[\frac{\sinh(\lambda_m^s X)}{\sinh(\lambda_m^s)} - \frac{\sin(\lambda_m^s X)}{\sin(\lambda_m^s)} \right],\qquad &\coth \lambda_m^s - \cot \lambda_m^s &= 0,\label{eq:efuncs_odd}\\[1.2mm]
    \psi^c_m(X) &= \frac{1}{\sqrt{2}}\left[\frac{\cosh(\lambda_m^c X)}{\cosh(\lambda_m^c)} - \frac{\cos(\lambda_m^c X)}{\cos(\lambda_m^c)} \right],\qquad &\tanh \lambda_m^c + \tan \lambda_m^c &= 0.\label{eq:efuncs_even}
\end{empheq}\end{subequations}
Here, $\psi^s_m(-X)=-\psi^s_m(X)$ are the odd, ``sine'' (``s''), eigenfunctions, while $\psi^c_m(-X)=\psi^c_m(X)$ are the even, ``cosine'' (``c''), eigenfunctions.  The respective sets of eigenvalues $\lambda_m^s$ and $\lambda_m^c$ are a countable ($m=1,2,\hdots$), real set of numbers, and satisfy the transcendental relations given in Eqs.~\eqref{eq:efuncs}.

As shown in \citep{papanicolaou2003,papanicolaou2009} (and the references therein, in particular \cite[\S133]{chandrasekhar1981}), the eigenfunctions $\psi_m^s$ and $\psi_m^c$ \emph{each} are orthonormal (for any $m,n=1,2,\hdots$):
\begin{subequations}%
\begin{align}
    \int_{-1}^{+1} \psi^s_m(X)\psi^s_{n}(X) \,dX &= \delta_{m,n},\\
    \int_{-1}^{+1} \psi^c_m(X)\psi^c_{n}(X) \,dX &= \delta_{m,n},
\end{align}
where $\delta_{m,n}$ is the Kronecker delta symbol, and \emph{mutually} orthogonal (for any $m=1,2,\hdots$):
\begin{equation}
    \int_{-1}^{+1} \psi^s_m(X)\psi^c_m(X) \,dX = 0.
\end{equation}
\end{subequations}%

Then, by standard theorems of ordinary differential equations \cite{coddington1955}, it follows that the eigenfunctions~\eqref{eq:efuncs} form a complete set in the space of square-integrable functions, $L^2[-1,+1]$, under the standard inner product. Thus, any function $U(X)\in L^2[-1,+1]$, has the following expansion:
\begin{subequations}\label{eq:efunc_expansion_w_coeff}\begin{align}
    U(X) &= \sum_{m=1}^\infty a_m^s \psi_m^s(X) + a_m^c \psi_m^c(X), \label{eq:efunc_expansion}\\
    a_m^s &\defeq \int_{-1}^{+1} U(X) \psi_m^s(X) \,dX,\\
    a_m^c &\defeq \int_{-1}^{+1} U(X) \psi_m^c(X) \,dX.
\end{align}\end{subequations}%
For the unsteady beam problem~\eqref{eq:beam_ibvp}, $U=U(X,T)$, and the coefficients may depend on time as $a_m^s=a_m^s(t)$ and $a_m^c=a_m^c(t)$.


\section{Peridynamic beam problem formulation}
\label{sec:pd_beam}

Under the theory of peridynamics \cite{silling2000,silling2010}, \citet{yang2022a} showed that the classical beam equation~\eqref{eq:dim_beam_eq} is generalized as the integrodifferential equation:
\begin{multline}
    m_s \frac{\partial^2 u}{\partial t^2} - \frac{B}{\delta^2} \int_{-\delta}^{+\delta} \frac{1}{\xi^2} \left[\int_{-\delta}^{+\delta} \frac{u(x+\eta)-u(x)}{\eta^2} \,d\eta \right.\\
    \left. - \int_{-\delta}^{+\delta} \frac{u(x+\xi+\eta)-u(x+\xi)}{\eta^2} \,d\eta \right] \,d\xi = p(x,t).
    \label{eq:pdbeam}
\end{multline}
The new feature of this equation is the \emph{horizon} size $\delta \ll \ell$. The horizon refers to the finite domain of influence for any material point $x$, which is the nonlocal feature of the theory. A larger $\delta$ value defines a bigger neighborhood around the material point $x$ that influences the behavior there, introducing interactions with more distant material points. The horizon size should be chosen based on the nature of the specific problem being addressed \cite{silling2000,silling2010}, for example, a nonzero $\delta$ should be used to study a nonlocal beam.

Again, introducing dimensionless variables based on the typical load scale $p_0$,
\begin{multline}
    T = \frac{t}{\sqrt{m_s\ell^4/B}},\qquad X = \frac{x}{\ell},\qquad \Xi = \frac{\xi}{\ell},\qquad H = \frac{\eta}{\ell},\qquad \Delta = \frac{\delta}{\ell},\\
    U(X,T) = \frac{u(x,t)}{p_0\ell^4/B},\qquad P(X,T) = \frac{p(x,t)}{p_0},
    \label{eq:nd_vars_pd}
\end{multline}
Eq.~\eqref{eq:pdbeam} becomes
\begin{multline}
    \frac{\partial^2 U}{\partial T^2} - \frac{1}{\Delta^2}\int_{-\Delta}^{+\Delta} \frac{1}{\Xi^2} \left[\int_{-\Delta}^{+\Delta} \frac{U(X+H)-U(X)}{H^2} \,dH \right.\\
    \left. - \int_{-\Delta}^{+\Delta} \frac{U(X+\Xi+H)-U(X+\Xi)}{H^2} \,dH \right] \,d\Xi = P(X,T) .
    \label{eq:pdbeam_nd}
\end{multline}
The IBVP comprised of Eqs.~\eqref{eq:pdbeam_nd}, \eqref{eq:clamped_bc}, and \eqref{eq:ic} is termed ``peridynamic beam theory'' (PBT) in this work.

Henceforth, we consider steady problems, so $U=U(X)$ and $P=P(X)$ only, and the governing equation is
\begin{multline}
    \frac{1}{\Delta^2}\int_{-\Delta}^{+\Delta} \frac{1}{\Xi^2} \left[\int_{-\Delta}^{+\Delta} \frac{U(X+H)-U(X)}{H^2} \,dH \right.\\
    \left. - \int_{-\Delta}^{+\Delta} \frac{U(X+\Xi+H)-U(X+\Xi)}{H^2} \,dH \right] \,d\Xi = -P(X).
    \label{eq:Yang}
\end{multline}
For sufficiently smooth $U(X)$, it may be verified by Taylor series expansions that the left-hand side of Eq.~\eqref{eq:Yang} reduces $-\partial^4 U /\partial X^4$ as $\Delta \to 0^+$, consistent with CTB~\eqref{eq:nd_beam_eq}.


\section{New solutions for clamped--clamped peridynamic beams}
\label{sec:solutions}

In this section, we show how Eq.~\eqref{eq:Yang} can be solved via a series of the form given in Eq.~\eqref{eq:efunc_expansion_w_coeff}.

\subsection{Odd solutions}
\label{sec:odd}

Consider an odd load such that $P(-X)=-P(X)$. Then, from the structure of the boundary-value problem, it is expected that $U(-X)=-U(X)$ is also an odd function. Thus, $a^c_m=0$ $\forall m$, and from Eq.~\eqref{eq:efunc_expansion} the solution can be expressed as 
\begin{equation}
    U(X) = \sum_{m=1}^\infty a_m^s \psi_m^s(X).
    \label{eq:oddseries}
\end{equation}
Next, substituting Eq.~\eqref{eq:oddseries} into Eq.~\eqref{eq:Yang}, we obtain
\begin{equation}
    \frac{1}{\sqrt{2}} \sum_{m=1}^\infty a_m^s   \left\{ \frac{\sinh(\lambda_m^sX)}{\sinh(\lambda_m^s)} \big[ \mathcal{I}_{1,m}^s(\Delta) \big]^2  - \frac{\sin(\lambda_m^sX)}{\sin(\lambda_m^s)} \big[ \mathcal{I}_{2,m}^s(\Delta) \big]^2 \right\} = P(X).
\label{eq:13}
\end{equation}
The full details of the calculation, including the definitions of $\mathcal{I}_{1,m}^s$ and $\mathcal{I}_{2,m}^s$, are given in the Appendix. 
Then, multiplying both sides of Eq.~\eqref{eq:13} by $\psi_n^s(X)$ and integrating over the domain yields
\begin{equation}
    \sum_{m=1}^\infty A^s_{n,m} a^s_m = P^s_n, \qquad n = 1,2,\hdots,
    \label{eq:14}
\end{equation}
where
\begin{multline}
    A^s_{n,m} \defeq \frac{1}{\sqrt{2}} \int_{-1}^{+1} \left\{ \frac{\sinh(\lambda_m^sX)}{\sinh(\lambda_m^s)} \big[ \mathcal{I}_{1,m}^s(\Delta) \big]^2 
    -\frac{\sin(\lambda_m^sX)}{\sin(\lambda_m^s)} \big[ \mathcal{I}_{2,m}^s(\Delta) \big]^2 \right\} \\
    \times \left[\frac{\sinh(\lambda_n^s X)}{\sinh(\lambda_n^s)} - \frac{\sin(\lambda_n^s X)}{\sin(\lambda_n^s)} \right] dX,
    \label{eq:Asnm}
\end{multline}
and
\begin{equation}
    P^s_n \defeq \int_{-1}^{+1} P(X) \left[\frac{\sinh(\lambda_n^s X)}{\sinh(\lambda_n^s)} - \frac{\sin(\lambda_n^s X)}{\sin(\lambda_n^s)} \right] dX.
\end{equation}
The integral in Eq.~\eqref{eq:Asnm} is evaluated analytically and given in the Appendix, Eq.~\eqref{eq:Asnm_evaluated}.

\begin{figure}[t]
    \centering
    \includegraphics[width=.9\textwidth]{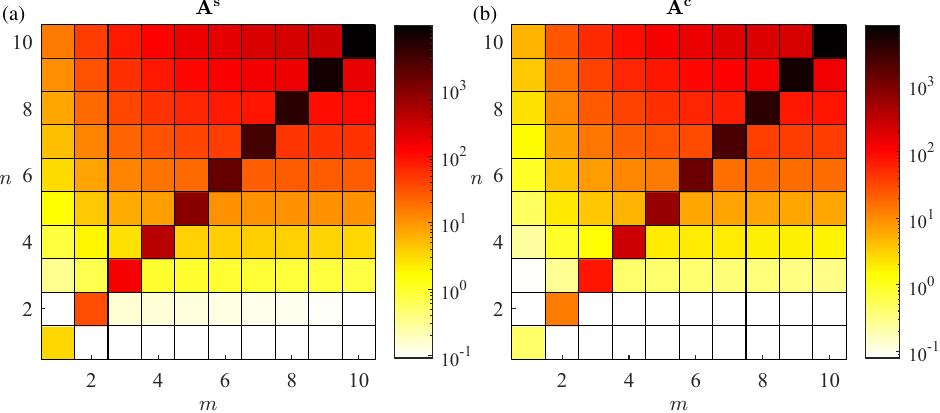}
    \caption{Visualization of matrices  (a) $\mathbf{A^s}$ and (b) $\mathbf{A^c}$, showing the magnitude of their entries, for a truncation at $M=10$ terms with $\Delta=0.1$.}
    \label{fig:matrices}
\end{figure}

The series in Eq.~\eqref{eq:oddseries} and the summation in Eq.~\eqref{eq:14} can be truncated at $M$ terms, leading to a square system of linear equations for the odd series coefficients $\{a^s_m\}_{m=1}^M$:
\begin{equation}
    \begin{pmatrix}
    A^s_{1,1} & A^s_{1,2} & \cdots & A^s_{1,M} \\
    A^s_{2,1} & A^s_{2,2} & \cdots & A^s_{2,M} \\
    \vdots  & \vdots  & \ddots & \vdots  \\
    A^s_{M,1} & A^s_{M,2} & \cdots & A^s_{M,M} 
    \end{pmatrix}
    \begin{pmatrix}
    a^s_1 \\ a^s_2  \\
    \vdots \\ a^s_M 
    \end{pmatrix} = \begin{pmatrix}
    P^s_1 \\ P^s_2  \\
    \vdots \\ P^s_M 
    \end{pmatrix},
    \label{eq:sinesys}
\end{equation}
where the square matrix, denoted by $\mathbf{A^s}$, is  non-symmetric but diagonally dominant as shown in Fig.~\ref{fig:matrices}(a) for $\Delta = 0.1$.

\begin{figure}
    \centering
    \includegraphics[width=1.0\textwidth]{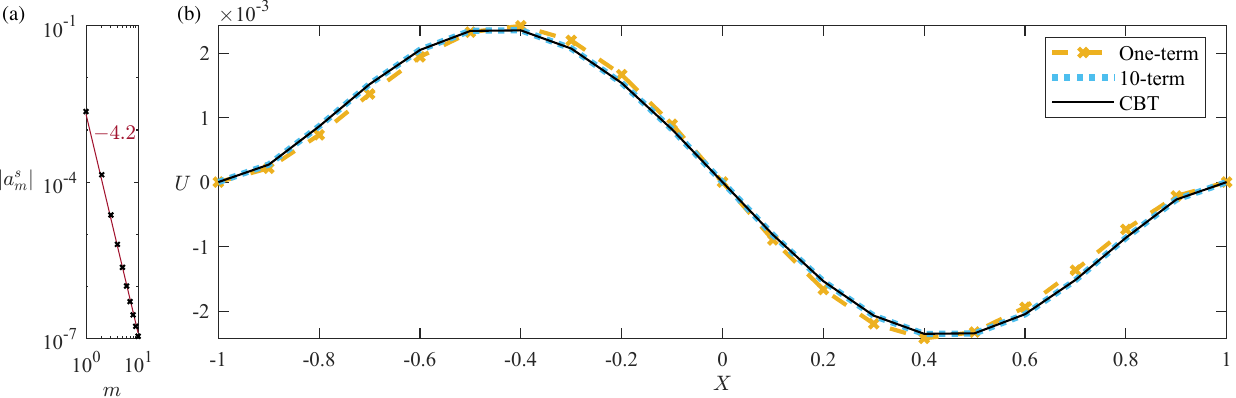}
    \caption{Beam bending under the load $P(X) = -X$, which allows using only odd eigenfunctions. (a) The absolute values of odd series coefficients, $|a_m^s|$. (b) Peridynamic beam theory (PBT) deflections calculated using a one-term truncation of the series~\eqref{eq:oddseries}, a ten-term truncation of the series~\eqref{eq:oddseries}, for $\Delta = 0.1$, and the classical beam theory (CBT) solution~\eqref{eq:CBT_odd_solution}.}
    \label{fig:px}
\end{figure}

Consider the example load, $P(X) = -X$, in which case
\begin{equation}
    P_n^s = -\frac{2}{(\lambda_n^s)^2} \big[\lambda_n^s\cot(\lambda_n^s)+\lambda_n^s\coth(\lambda_n^s)-2\big].
\end{equation}
The beam deflection can be represented as an odd eigenfunction series. The absolute values of the first ten expansion coefficients are plotted in Fig.~\ref{fig:px}(a), alongside a reference line of slope $-4.2$. The coefficients' rapid decay at a rate $\mathcal{O}(m^{-4})$ highlights the efficiency of using fourth-order eigenfunction series to solve for the deflection under PBT, as the series~\eqref{eq:oddseries} converges quite rapidly.

In Fig.~\ref{fig:px}(b), we compare the beam deflection computed using only the first term with that calculated using the first ten terms of the series~\eqref{eq:oddseries} and the deflection from the attendant CBT solution, i.e., the solution of the steady BVP~\eqref{eq:beam_ibvp} with the same load $P(X) = -X$, which is easily found to be:
\begin{equation}
    U_\mathrm{CBT}(X) = -\frac{1}{120}\big(X^5-2X^3+X\big).
    \label{eq:CBT_odd_solution}
\end{equation}

From Fig.~\ref{fig:px}(b), it can be seen that, surprisingly, a single-term approximation can provide a satisfactory solution. For the value $\Delta = 0.1$ of the horizon chosen in this figure, the difference between the CBT and the PBT is small.

\subsection{Even solutions}
\label{sec:even}

Consider an even load such that $P(-X)=P(X)$. Again, from the structure of the boundary-value problem, it is expected that $U(-X)=U(X)$ is also an even function. Thus, $a^s_m=0$ $\forall m$, and from Eq.~\eqref{eq:efunc_expansion} the solution can be expressed as 
\begin{equation}
    U(X) = \sum_{m=1}^\infty a_m^c \psi_m^c(X).
    \label{eq:evenseries}
\end{equation}
Next, substituting Eq.~\eqref{eq:evenseries} into Eq.~\eqref{eq:Yang}, we obtain
\begin{equation}
    \frac{1}{\sqrt{2}} \sum_{m=1}^\infty a_m^c  \left\{\frac{\cosh(\lambda_m^cX)}{\cosh(\lambda_m^c)} \big[ \mathcal{I}_{1,m}^c(\Delta) \big]^2 -\frac{\cos(\lambda_m^cX)}{\cos(\lambda_m^c)} \big[ \mathcal{I}_{2,m}^c(\Delta) \big]^2 \right\} =P(X).
    \label{eq:19}
\end{equation}
The full details of the calculation, including the definitions of $\mathcal{I}_{1,m}^c$ and $\mathcal{I}_{2,m}^c$, are given in the Appendix. Then, multiplying both sides of Eq.~\eqref{eq:19} by $\psi_n^c(X)$ and integrating over the domain yields
\begin{equation}
    \sum_{m=1}^\infty A^c_{n,m} a^c_m = P^c_n,\qquad n=1,2,\hdots,
    \label{eq:20}
\end{equation}
where
\begin{multline}
    A^c_{n,m} \defeq \frac{1}{\sqrt{2}} \int_{-1}^{+1} \left\{\frac{\cosh(\lambda_m^cX)}{\cosh(\lambda_m^c)} \big[ \mathcal{I}_{1,m}^c(\Delta) \big]^2  -\frac{\cos(\lambda_m^cX)}{\cos(\lambda_m^c)} \big[ \mathcal{I}_{2,m}^c(\Delta) \big]^2 \right\}  \\
    \times \left[\frac{\cosh(\lambda_n^c X)}{\cosh(\lambda_n^c)} - \frac{\cos(\lambda_n^c X)}{\cos(\lambda_n^c)} \right] dX,
    \label{eq:Acnm}
\end{multline}
and
\begin{equation}
    P^c_n \defeq \int_{-1}^{+1}  P(X) \left[ \frac{\cosh(\lambda_n^c X)}{\cosh(\lambda_n^c)} - \frac{\cos(\lambda_n^c X)}{\cos(\lambda_n^c)} \right] dX.
\end{equation}
The integral in Eq.~\eqref{eq:Acnm} is evaluated analytically and given in the Appendix, Eq.~\eqref{eq:Acnm_evaluated}.

As before, in Eq.~\eqref{eq:evenseries} and the summation in Eq.~\eqref{eq:20} can be truncated at $M$ terms, leading to a square system of linear equations for the even series coefficients $\{a^c_m\}_{m=1}^M$:
\begin{equation}
    \begin{pmatrix}
    A^c_{1,1} & A^c_{1,2} & \cdots & A^c_{1,M} \\
    A^c_{2,1} & A^c_{2,2} & \cdots & A^c_{2,M} \\
    \vdots  & \vdots  & \ddots & \vdots  \\
    A^c_{M,1} & A^c_{M,2} & \cdots & A^c_{M,M} 
    \end{pmatrix}
    \begin{pmatrix}
    a^c_1 \\ a^c_2  \\
    \vdots \\ a^c_M 
    \end{pmatrix} = \begin{pmatrix}
    P^c_1 \\ P^c_2  \\
    \vdots \\ P^c_M 
    \end{pmatrix},
    \label{eq:oddsys}
\end{equation}
where the square matrix, denoted by $\mathbf{A^c}$, is non-symmetric but diagonally dominant as shown in Fig.~\ref{fig:matrices}(b) for $\Delta = 0.1$.

\begin{figure}[b]
    \centering
    \includegraphics[width=1.0\textwidth]{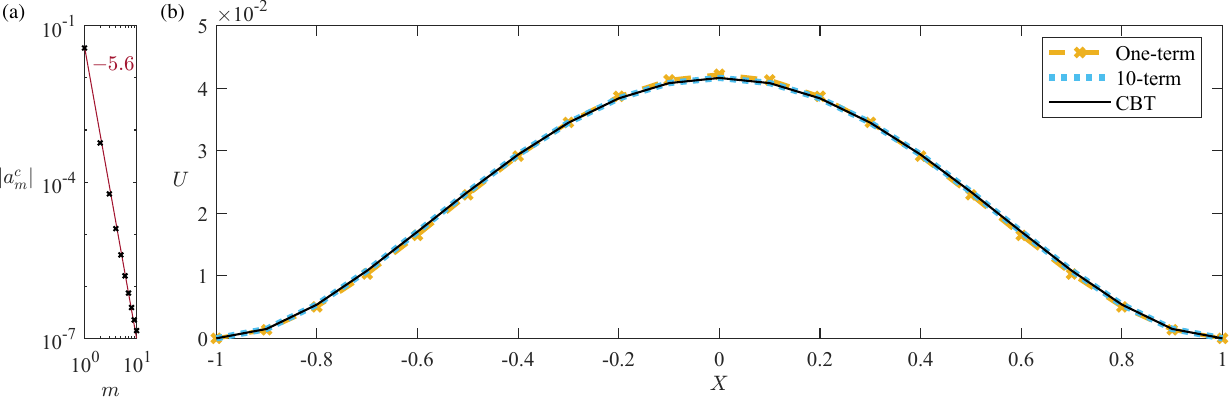}
    \caption{Beam bending under the uniform load $P(X) \equiv 1$, which allows using only even eigenfunctions. (a) The absolute values of even series coefficients, $|a_m^c|$. (b) Peridynamic beam theory (PBT) deflections calculated using a one-term truncation of the series~\eqref{eq:evenseries}, a ten-term truncation of the series~\eqref{eq:evenseries}, for $\Delta = 0.1$, and the classical beam theory (CBT) solution~\eqref{eq:CBT_even_solution}.}    
    \label{fig:pconst}
\end{figure}

Consider the example load, $P(X) \equiv 1$, in which case
\begin{equation}
    P_n^c = -\frac{2}{\lambda_n^c} \big[\tan(\lambda_n^c)-\tanh(\lambda_n^c)\big].
\end{equation}
The beam deflection can be represented as an even eigenfunction series~\eqref{eq:evenseries}. We calculated and plotted the absolute values of the first ten expansion coefficients in Fig.~\ref{fig:pconst}(a), alongside a reference line with slope of $-5.6$. The coefficients' rapid decay at a rate $\mathcal{O}(m^{-5})$, which is significantly faster than fourth order, again highlights the efficiency of using fourth-order eigenfunction series to solve for the deflection under PBT. 

In Fig.~\ref{fig:pconst}(b), we compare the PBT deflection computed using only the first term with that calculated using the first ten terms of the series~\eqref{eq:evenseries} and the deflection from the attendant CBT solution, i.e., the solution of the steady BVP~\eqref{eq:beam_ibvp} with the same load $P(X) \equiv 1$, which is easily found to be:
\begin{equation}
    U_\mathrm{CBT}(X) = \frac{1}{24}(X-1)^2(X+1)^2.
    \label{eq:CBT_even_solution}
\end{equation}

Once again, it can be seen from Fig.~\ref{fig:pconst}(b) that a single-term approximation based on the fourth-order eigenfunctions can provide a satisfactory solution.

\subsection{General case}
\label{sec:general}

As the third example, consider the deflection of a beam caused by a step load, represented as the Heaviside unit step function: $P(X) = H(X)$. In this case,
\begin{subequations}\begin{align}
    P_n^s &= \frac{1}{\lambda_n^s} \big[\cot(\lambda_n^s)-\csc(\lambda_n^s)+\tanh(\lambda_n^s/2)\big],\\
    P_n^c &= -\frac{1}{\lambda_n^c} \big[\tan(\lambda_n^c)-\tanh(\lambda_n^c)\big].
\end{align}\end{subequations}
Now, the solution $U(X)$ is neither odd nor even, and \emph{both} the odd and even series coefficients have to be computed. Hence, $U(X)$ is expressed as in Eq.~\eqref{eq:efunc_expansion}. Substituting Eq.~\eqref{eq:efunc_expansion} into Eq.~\eqref{eq:Yang}, we obtain
\begin{multline}
    \frac{1}{\sqrt{2}} \sum_{m=1}^\infty a_m^s   \left\{ \frac{\sinh(\lambda_m^sX)}{\sinh(\lambda_m^s)} \big[ \mathcal{I}_{1,m}^s(\Delta) \big]^2  - \frac{\sin(\lambda_m^sX)}{\sin(\lambda_m^s)} \big[ \mathcal{I}_{2,m}^s(\Delta) \big]^2 \right\}\\
     + \frac{1}{\sqrt{2}} \sum_{m=1}^\infty a_m^c  \left\{\frac{\cosh(\lambda_m^cX)}{\cosh(\lambda_m^c)} \big[ \mathcal{I}_{1,m}^c(\Delta) \big]^2 -\frac{\cos(\lambda_m^cX)}{\cos(\lambda_m^c)} \big[ \mathcal{I}_{2,m}^c(\Delta) \big]^2 \right\}= P(X).
     \label{eq:subfull}
\end{multline}
Multiplying both sides of Eq.~\eqref{eq:subfull} by $\psi_n^s(X)$ and integrating over the domain yields 
\begin{equation}
    \sum_{m=1}^\infty A^s_{n,m} a^s_m = P^s_n, \qquad n = 1,2,\hdots,
\end{equation}
where $A^s_{n,m}$ is given in Eq.~\eqref{eq:Asnm_evaluated} in the Appendix. Likewise, multiplying both sides of Eq.~\eqref{eq:subfull} by $\psi_n^c(X)$ and integrating over the domain yields 
\begin{equation}
    \sum_{m=1}^\infty A^c_{n,m} a^c_m = P^c_n,\qquad n=1,2,\hdots,
\end{equation}
where $A^c_{n,m}$ is given in Eq.~\eqref{eq:Acnm_evaluated} in the Appendix. Note that, upon truncating the series in this case, we have to solve \emph{two} linear matrix equations, one for the odd coefficients and another for the even coefficients.

\begin{figure}[t]
    \centering
    \includegraphics[width=1.0\textwidth]{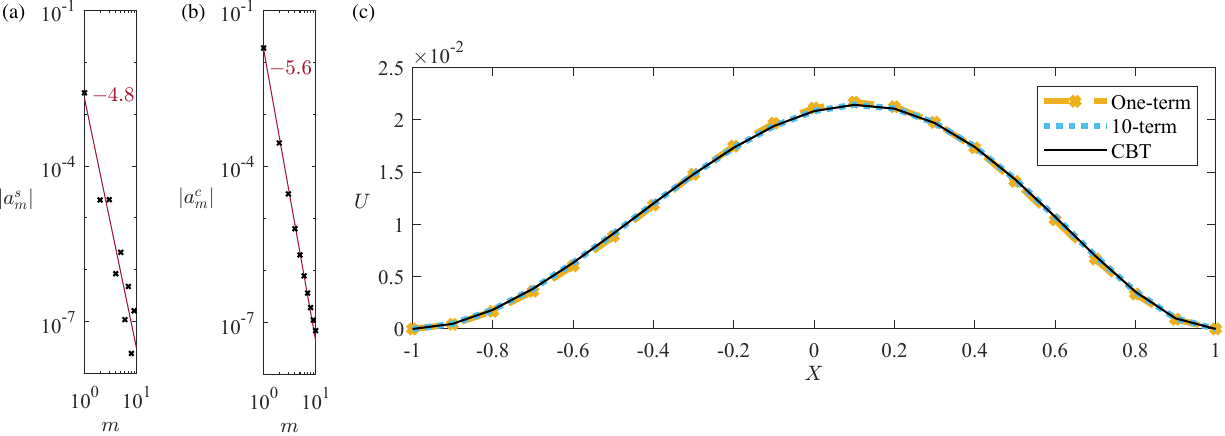}
    \caption{Beam bending under the step load $P(X) = H(X)$, using both even and odd fourth-order eigenfunctions. (a) The absolute values of odd coefficients, $| a_m^s |$. (b) The absolute values of even coefficients, $| a_m^c |$. (c) Peridynamic beam theory (PBT) deflections calculated using a one-term truncation of the series~\eqref{eq:efunc_expansion}, a ten-term truncation of the series~\eqref{eq:efunc_expansion}, for $\Delta = 0.002$, and the classical beam theory (CBT) solution~\eqref{eq:CBT_step_solution}.}    
    \label{fig:pstep}
\end{figure}

The absolute values of the odd and even coefficients are plotted in Fig.~\ref{fig:pstep}(a) and in Fig.~\ref{fig:pstep}(b), respectively. As before, the odd coefficients exhibit a decay rate of $\mathcal{O}(m^{-4})$, whereas the even coefficients diminish at a rate of $\mathcal{O}(m^{-5})$, as shown by the reference lines of slope $-4.8$ and $-5.6$, respectively.
The closed-form analytical solution for the beam deflection based on CBT, i.e., the solution of the steady BVP~\eqref{eq:beam_ibvp} with the same load $P(X)=H(X)$, is easily found as:
\begin{equation}
    U_\mathrm{CBT}(X) = \frac{1}{96}\big[4H(X)X^3-3X^2-4X+3\big].
    \label{eq:CBT_step_solution}
\end{equation}

In Fig.~\ref{fig:pstep}(c), we compare the beam deflection computed using only the first term (of both the even and odd functions, so two terms total) with that calculated using the first ten terms of the series (of both the even and odd functions, so twenty terms total) and the deflection from the attendant CBT solution. Once again, it can be seen that a single-term approximation can yield a satisfactory solution.

\subsection{Comparison to the Fourier sine series solution}

\begin{figure}[t]
    \centering
    \includegraphics[width=1.0\textwidth]{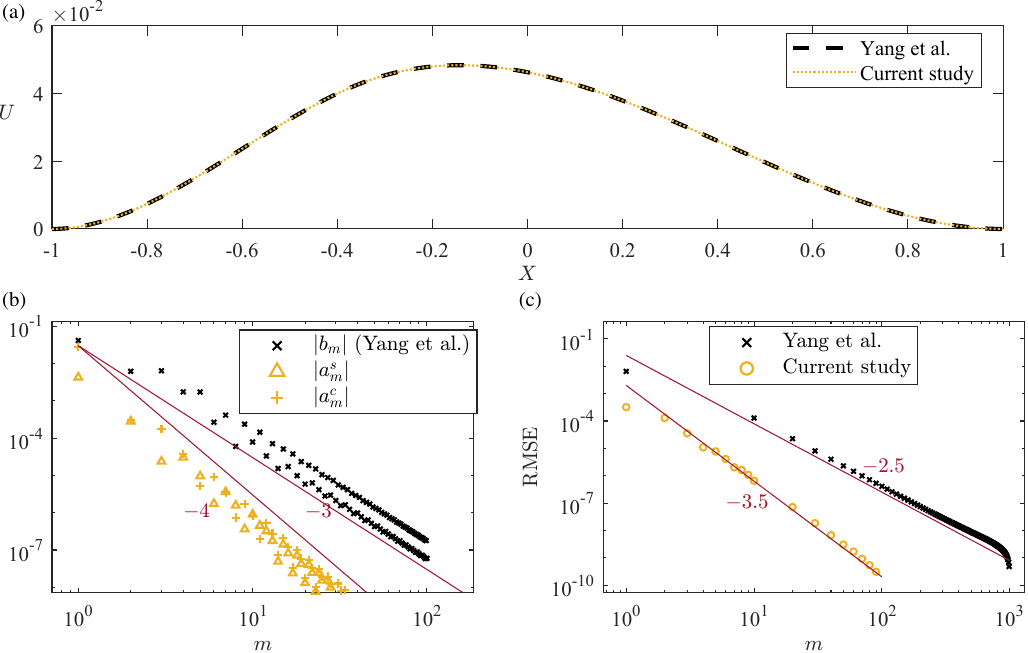}
    \caption{(a) Comparison of different solutions for the peridynamic beam theory (PBT) deflection under the point-load $P(X)=(3/2)\delta(3X+1)$: the sine series of \citet{yang2023a} (dashed) and beam function series of this work (dotted). (b) The absolute values of the series' coefficients from each approach. (c) The root-mean-squared error (RMSE) as a function of the number of terms $m$ in the truncated sum for each series.}
    \label{fig:Yang}
\end{figure}

\citet{yang2023a} employed a Fourier sine series to determine the deflection of a peridynamic beam~\eqref{eq:pdbeam_nd} subjected to the point load $P(X) = (3/2)\delta(3X+1)$, where here $\delta(\cdot)$ is the Dirac delta, \emph{not} the dimensional horizon. In this case,
\begin{subequations}\begin{align}
    P_n^s &= \frac{3}{2} \left[ \frac{\sin(\lambda_n^s/3)}{\sin(\lambda_n^s)}-\frac{\sinh(\lambda_n^s/3)}{\sinh(\lambda_n^s)} \right],\\
    P_n^c &= \frac{3}{2} \left[ \frac{\cosh(\lambda_n^c/3)}{\cosh(\lambda_n^c)}-\frac{\cos(\lambda_n^c/3)}{\cos(\lambda_n^c)} \right].
\end{align}\end{subequations}
We made the solution from \cite{yang2023a} dimensionless using our variables from Eq.~\eqref{eq:nd_vars_pd} and transformed it to the domain $X\in[-1,+1]$. To align with the setup of \citet{yang2023a} (Sec.~5.1 therein), we take $\Delta = 0.002$. In Fig.~\ref{fig:Yang}(a), Yang et al.'s solution, truncated at $1000$ terms, is plotted alongside our solution based on the fourth-order eigenfunctions series~\eqref{eq:efunc_expansion_w_coeff}, truncated at $10$ terms. Clearly, the two series solutions agree well. 

The decay rate of the absolute values of the Fourier sine series' coefficients is $\mathcal{O}(m^{-3})$, as illustrated by a reference line in Fig.~\ref{fig:Yang}(b). In contrast,  $|a_m^s|$ and $|a_m^c|$ from our study, also plotted in Fig.~\ref{fig:Yang}(b), have a convergence rate of $\mathcal{O}(m^{-4})$. Notably, while the Fourier series' coefficients decrease to $10^{-7}$ after $100$ terms, the fourth-order eigenfunction series' coefficients achieve this magnitude after just $10$ terms. 

Furthermore, we calculated the root mean square error (RMSE) between the $m$-term sum truncation and $1000$-term sum truncation for both Yang et al.'s and our series solutions. These results are shown in Fig.~\ref{fig:Yang}(c). The reference lines indicate that the sine series' partial sums converge at a rate $\sim m^{-2.5}$, whereas the fourth-order eigenfunction series' partial sums converge one order faster at a rate $\sim m^{-3.5}$. This observation means the proposed fourth-order eigenfunction series solution~\eqref{eq:efunc_expansion_w_coeff} can achieve a comparable precision with far fewer terms (roughly $10$ times fewer for large $m$).

\begin{figure}[t]
    \centering
    \includegraphics[width=0.9\textwidth]{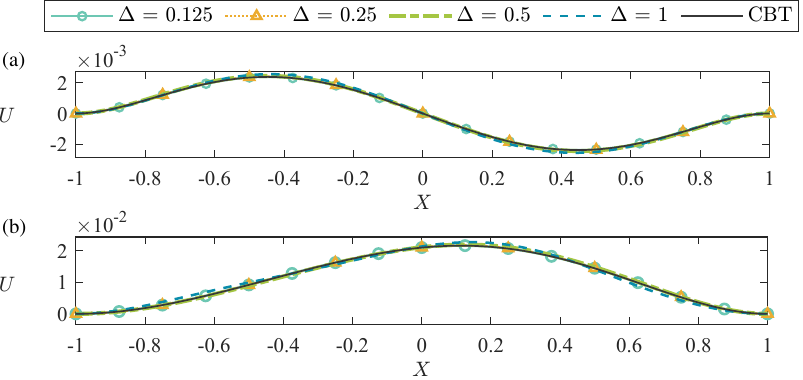}
    \caption{The effect of varying the horizon on the deflection under the peridynamic beam theory (PBT), computed as a beam function series and truncated at $5$ terms, and as compared to the corresponding solution under the classical beam theory (CBT), under the loads (a) $P(X) = -X$ and (b) $P(X) = H(X)$.}
    \label{fig:delta}
\end{figure}

\subsection{Effect of the peridynamic horizon}

Finally, we investigate the influence of the horizon size $\Delta$ on the solutions involving both a continuous and a discontinuous load, specifically $P(X) = -X$ and $P(X) = H(X)$, from Sec.~\ref{sec:even} and Sec.~\ref{sec:general}, respectively. The results are shown in Fig.~\ref{fig:delta}(a) and Fig.~\ref{fig:delta}(b), respectively. To provide a benchmark, the analytical solutions obtained from CBT, namely Eqs.~\eqref{eq:CBT_odd_solution} and \eqref{eq:CBT_step_solution}, are plotted alongside for both cases. It is expected that as $\Delta \to 0^+$, the PBT should reduce to the CBT. Our analysis indeed reveals that diminishing $\Delta$ leads to convergence of the peridynamic solution to the classical one. Remarkably, even a relatively ``large'' value of $\Delta=1$ provides a solution that looks similar to the CBT one, though some differences are apparent on zooming in. While it may be that such a large $\Delta$ value is unrealistic for some material being described by peridynamic theory, here we are just performing a mathematical exercise and exploring the outcomes.


\section{Conclusion}

In this work, we proposed a new analytical approach for solving for the displacement of clamped--clamped nonlocal generalizations of the Euler--Bernoulli beam under a peridynamic theory. Specifically, in Sec.~\ref{sec:beam_efuncs}, we introduced the ``natural'' fourth-order eigenfunctions that satisfy the clamped--clamped boundary conditions arising from the classical beam theory. Using these eigenfunctions, we derived solutions for static bending of the peridynamic beam introduced in Sec.~\ref{sec:pd_beam} subject to several representative loads --- including ones that engage only the ``odd'' or ``even'' eigenfunctions, as well as both.

In Sec.~\ref{sec:solutions}, we compared our series solutions based on the fourth-order eigenfunctions to those previously obtained as Fourier sine series. The series solutions based on fourth-order eigenfunctions converge much faster than the Fourier sine series. Specifically, for an example of an offset point load on the beam, we showed that the expansion coefficients of the series based on fourth-order eigenfunctions decay as the number of terms to the fourth power, as opposed to the third power for the sine series. Further, the partial sums of the series based on the fourth-order eigenfunction were found to decay one order faster than the sine series. Thus, despite the fact that the fourth-order eigenfunctions do not diagonalize the peridynamic operator (like sines do), the matrix that has to be inverted to obtain the coefficients of an accurate series solution is small.

The peridynamic model discussed in this work is essentially linear but nonlocal. An interesting generalization would be to consider \emph{nonlinear} nonlocal interactions with a horizon, perhaps along the lines of the idea introduced by \citet{porubov2018} within the context of lattice dynamics. 
In future work, it would also be of interest to couple the peridynamic beam equation~\eqref{eq:pdbeam} with the flow of a thin film underneath it (a fluid--structure interaction problem), which may also lead to an IBVP that can be solved analytically using eigenfunction expansions \cite{NectarIvan2023}.

\begin{acknowledgement}
This research was supported by the U.S.\ National Science Foundation under grant No.\ CMMI-2245343. We dedicate this work to the 60\textsuperscript{th} birthday of Prof.\ Alexey V.\ Porubov. We also thank Prof.\ Victor A.\ Eremeyev for his kind invitation to contribute to this volume and for his efforts in co-organizing it. I.C.C.\ further acknowledges many insightful discussions with Prof.\ Nectarios C.\ Papanicolaou on beam eigenfunctions.
\end{acknowledgement}

\ethics{Competing Interests}{The authors have no conflicts of interest to declare that are relevant to the content of this chapter.}


\section*{Appendix}

In this Appendix, we provide the formulas for evaluating the integrals in the peridynamic beam equation~\eqref{eq:Yang}, when the displacement $U$ is expanded in a series~\eqref{eq:efunc_expansion_w_coeff} in terms of the fourth-order eigenfunctions~\eqref{eq:efuncs}.

First, for the case of the odd function expansion in Eq.~\eqref{eq:oddseries}, we have
\begin{subequations}
\begin{multline}
    \frac{1}{\Delta}\int_{-\Delta}^{+\Delta} \frac{U(X+H)-U(X)}{H^2} \,dH \\
    = \frac{1}{\sqrt{2}} \sum_{m=1}^\infty a_m^s \left[\frac{\sinh(\lambda_m^sX)}{\sinh(\lambda_m^s)} \mathcal{I}_{1,m}^s(\Delta) - \frac{\sin(\lambda_m^sX)}{\sin(\lambda_m^s)} \mathcal{I}_{2,m}^s(\Delta) \right] ,
\end{multline}
and
\begin{multline}
    \frac{1}{\Delta}\int_{-\Delta}^{+\Delta} \frac{U(X+\Xi+H)-U(X+\Xi)}{H^2} \,dH \\
    = \frac{1}{\sqrt{2}} \sum_{m=1}^\infty a_m^s \left[\frac{\sinh\big(\lambda_m^s(X+\Xi)\big)}{\sinh(\lambda_m^s)} \mathcal{I}_{1,n}^s(\Delta) - \frac{\sin\big(\lambda_m^s(X+\Xi)\big)}{\sin(\lambda_m^s)} \mathcal{I}_{2,m}^s(\Delta) \right] ,
\end{multline}
\end{subequations}
where
\begin{subequations}\label{eq:I12s}\begin{align}
    \mathcal{I}_{1,m}^s(\Delta) &\defeq \frac{1}{\Delta}\int_{-\Delta}^{+\Delta}\frac{1}{H^2}\big[\cosh(\lambda_m^sH)-1\big] \,dH \\
    &=\frac{2}{\Delta^2} \big[ \Delta\lambda_m^s \Shi(\Delta\lambda_m^s)-\cosh(\Delta\lambda_m^s)+1 \big] \to (\lambda_m^s)^2 \quad\text{as}\quad \Delta \to 0^+,\nonumber \\
    \mathcal{I}_{2,m}^s(\Delta) &\defeq \frac{1}{\Delta}\int_{-\Delta}^{+\Delta}\frac{1}{H^2}\big[\cos(\lambda_m^sH)-1\big] \,dH \\
    &=-\frac{2}{\Delta^2} \big[ \Delta\lambda_m^s \Si(\Delta\lambda_m^s)+\cos(\Delta\lambda_m^s)-1 \big] \to - (\lambda_m^s)^2 \quad\text{as}\quad \Delta \to 0^+,\nonumber
\end{align}\end{subequations}
and $\Si(z) \defeq \int_0^z \zeta^{-1} \sin\zeta \,d\zeta$ and $\Shi(z) \defeq \int_0^z \zeta^{-1} \sinh\zeta \,d\zeta$ are the sine and hyperbolic sine integrals, respectively \cite[\S6.2]{DLMF}.

The integral definining $A^s_{n,m}$ in Eq.~\eqref{eq:Asnm} can be evaluated to yield
\begin{equation}
    A^s_{n,m} = \frac{1}{\sqrt{2}} \begin{cases} \big[\mathcal{I}_{1,m}^s(\Delta)\big]^2 \big[ \coth(\lambda_m^s)/\lambda_m^s + 1 - \coth^2(\lambda_m^s) \big] \\
    \qquad +\big[\mathcal{I}_{2,m}^s(\Delta)\big]^2 \big[ -\cot(\lambda_m^s)/\lambda_m^s + 1 + \cot^2(\lambda_m^s) \big], & \hspace{-5mm} \text{if }n=m,\\[2mm]
    2 \big[\mathcal{I}_{1,m}^s(\Delta)\big]^2 \big\{ \big[\lambda_n^s\coth(\lambda_n^s)-\lambda_m^s\coth(\lambda_m^s)\big]/\big[(\lambda_n^s)^2-(\lambda_m^s)^2\big] \\
    \qquad -\big[\lambda_m^s\coth(\lambda_m^s)-\lambda_n^s\cot(\lambda_n^s)\big]/({\lambda_n^s}^2+{\lambda_m^s}^2)\big\} \\
    + 2 \big[\mathcal{I}_{2,m}^s(\Delta)\big]^2 \big\{ \big[\lambda_m^s\cot(\lambda_m^s)-\lambda_n^s\cot(\lambda_n^s)\big]/\big[(\lambda_n^s)^2-(\lambda_m^s)^2\big] \\
    \qquad -\big[\lambda_n^s\coth(\lambda_n^s)-\lambda_m^s\cot(\lambda_m^s)\big]/\big[(\lambda_n^s)^2+(\lambda_m^s)^2\big] \big\}, & \hspace{-5mm} \text{if }n\ne m.
\end{cases}
\label{eq:Asnm_evaluated}
\end{equation}
From the definitions in Eq.~\eqref{eq:I12s} and the eigenvalue relation in Eq.~\eqref{eq:efuncs_odd}, it is easy to see that $A^s_{n,m} = \sqrt{2} (\lambda_m^s)^4 \delta_{n,m}$ as $\Delta \to 0^+$, as expected, noting that we multiplied through by $\sqrt{2}$ in Eq.~\eqref{eq:14}.

Second, for the case of the even function expansion in Eq.~\eqref{eq:evenseries}, we have
\begin{subequations}
\begin{multline}
    \frac{1}{\Delta}\int_{-\Delta}^{+\Delta} \frac{U(X+H)-U(X)}{H^2} \,dH \\
    = \frac{1}{\sqrt{2}} \sum_{m=1}^\infty a_m^c \left[\frac{\cosh(\lambda_m^cX)}{\cosh(\lambda_m^c)} \mathcal{I}_{1,m}^c(\Delta)  - \frac{\cos(\lambda_m^cX)}{\cos(\lambda_m^c)} \mathcal{I}_{2,m}^c(\Delta) \right],
\end{multline}
and
\begin{multline}
    \frac{1}{\Delta}\int_{-\Delta}^{+\Delta} \frac{U(X+\Xi+H)-U(X+\Xi)}{H^2} \,dH \\ 
    = \frac{1}{\sqrt{2}} \sum_{m=1}^\infty a_m^c \left[\frac{\cosh\big(\lambda_m^c(X+\Xi)\big)}{\cosh(\lambda_m^c)} \mathcal{I}_{1,m}^c(\Delta) - \frac{\cos\big(\lambda_m^c(X+\Xi)\big)}{\cos(\lambda_m^c)} \mathcal{I}_{2,m}^c(\Delta) \right] ,
\end{multline}
\end{subequations}
where
\begin{subequations}\label{eq:I12c}\begin{align}
    \mathcal{I}_{1,m}^c(\Delta) &\defeq \frac{1}{\Delta}\int_{-\Delta}^{+\Delta}\frac{1}{H^2}\big[\cosh(\lambda_m^cH)-1\big] \,dH ,\\
    &=\frac{2}{\Delta^2} \big[ \Delta\lambda_m^c \Shi(\Delta\lambda_m^c)-\cosh(\Delta\lambda_m^c)+1 \big] \to (\lambda_m^c)^2 \quad\text{as}\quad \Delta \to 0^+,\nonumber\\
    \mathcal{I}_{2,m}^c(\Delta) &\defeq \frac{1}{\Delta}\int_{-\Delta}^{+\Delta}\frac{1}{H^2}\big[\cos(\lambda_m^cH)-1\big] \,dH \\
    &=-\frac{2}{\Delta^2} \big[ \Delta\lambda_m^c \Si(\Delta\lambda_m^c)+\cos(\Delta\lambda_m^c)-1 \big] \to -(\lambda_m^c)^2 \quad\text{as}\quad \Delta \to 0^+.\nonumber
\end{align}\end{subequations}

The integral definining $A^c_{n,m}$ in Eq.~\eqref{eq:Acnm} can be evaluated to yield
\begin{equation}
    A^c_{n,m} = \frac{1}{\sqrt{2}} \begin{cases} \big[\mathcal{I}_{1,m}^c(\Delta)\big]^2 \big[ \tanh(\lambda_m^c)/\lambda_m^c + 1 - \tanh^2(\lambda_m^c) \big] \\ 
    \qquad + \big[\mathcal{I}_{2,m}^c(\Delta)\big]^2 \big[ \tan(\lambda_m^c)/\lambda_m^c + 1 +  \tan^2(\lambda_m^c) \big] , & \hspace{-10mm} \text{if }n=m ,\\[2mm] 
    2 \big[\mathcal{I}_{1,m}^c(\Delta)\big]^2 \big\{\big[\lambda_n^c\tanh(\lambda_n^c)-\lambda_m^c\tanh(\lambda_m^c)\big]/\big[(\lambda_n^c)^2-(\lambda_m^c)^2\big] \\
    \qquad - \big[\lambda_m^c\tanh(\lambda_m^c)+\lambda_n^c\tan(\lambda_n^c)\big]/\big[(\lambda_n^c)^2+(\lambda_m^c)^2\big] \big\} \\
    + 2 \big[\mathcal{I}_{2,m}^c(\Delta)\big]^2 \big\{\big[-\lambda_m^c\tan(\lambda_m^c)+\lambda_n^c\tan(\lambda_n^c)\big]/\big[(\lambda_n^c)^2-(\lambda_m^c)^2\big] \\
    \qquad -\big[\lambda_n^c\tanh(\lambda_n^c)+\lambda_m^c\tan(\lambda_m^c)\big]/\big[(\lambda_n^c)^2+(\lambda_m^c)^2\big]  \big\}, & \hspace{-10mm} \text{if }n\ne m  .
\end{cases}
\label{eq:Acnm_evaluated}
\end{equation}
From the definitions in Eq.~\eqref{eq:I12c} and the eigenvalues relation in Eq.~\eqref{eq:efuncs_even}, it is easy to see that $A^c_{n,m} = \sqrt{2} (\lambda_m^c)^4 \delta_{n,m}$ as $\Delta \to 0^+$, as expected, noting that we multiplied through by $\sqrt{2}$ in Eq.~\eqref{eq:20}.



\clearpage
\footnotesize{
\bibliographystyle{spbasic-mod}
\bibliography{references.bib}

\begin{thebibliography}{31}
\providecommand{\natexlab}[1]{#1}
\providecommand{\url}[1]{{#1}}
\providecommand{\urlprefix}{URL }
\expandafter\ifx\csname urlstyle\endcsname\relax
  \providecommand{\doi}{DOI~\discretionary{}{}{}\begingroup \urlstyle{rm}\Url}\else
  \providecommand{\doi}[1]{DOI~\discretionary{}{}{}\href{http://dx.doi.org/#1}{#1}}
\providecommand{\eprint}[2][]{\url{#2}}

\bibitem[{Altenbach and Zhilin(2004)}]{altenbach2004}
Altenbach H, Zhilin PA (2004) The theory of simple elastic shells. In: Kienzler R, Ott I, Altenbach H (eds) Theories of Plates and Shells: Critical Review and New Applications, Springer, Berlin/Heidelberg, pp 1--12. \doi{10.1007/978-3-540-39905-6\_1}

\bibitem[{Altenbach et~al.(2011)Altenbach, Maugin, and Erofeev}]{altenbach2011}
Altenbach H, Maugin GA, Erofeev V (eds)  (2011) Mechanics of Generalized Continua, Advanced Structured Materials, vol~7. Springer-Verlag, Berlin/Heidelberg. \doi{10.1007/978-3-642-19219-7}

\bibitem[{Challamel(2018)}]{challamel2018}
Challamel N (2018) Static and dynamic behaviour of nonlocal elastic bar using integral strain-based and peridynamic models. Comptes Rendus -- Mecanique \textbf{346}, 320--335. \doi{10.1016/j.crme.2017.12.014}

\bibitem[{Chandrasekhar(1981)}]{chandrasekhar1981}
Chandrasekhar S (1981) Hydrodynamic and Hydromagnetic Stability. Dover Publications, New York, NY

\bibitem[{Coddington and Levinson(1955)}]{coddington1955}
Coddington EA, Levinson N (1955) Theory of Ordinary Differential Equations. McGraw-Hill, New York, NY

\bibitem[{Diyaroglu et~al.(2015)Diyaroglu, Oterkus, Oterkus, and Madenci}]{diyaroglu2015}
Diyaroglu C, Oterkus E, Oterkus S, Madenci E (2015) Peridynamics for bending of beams and plates with transverse shear deformation. Int J Solids Struct \textbf{69-70}, 152--168. \doi{10.1016/j.ijsolstr.2015.04.040}

\bibitem[{{\relax DLMF}(2024)}]{DLMF}
{\relax DLMF} (2024) {\it NIST Digital Library of Mathematical Functions}. Release 1.2.0 of 2024-03-15. \urlprefix\url{https://dlmf.nist.gov/}, {F.~W.~J. Olver, A.~B. {Olde Daalhuis}, D.~W. Lozier, B.~I. Schneider, R.~F. Boisvert, C.~W. Clark, B.~R. Miller, B.~V. Saunders, H.~S. Cohl, and M.~A. McClain, eds.}

\bibitem[{Huang(2018)}]{huang2018}
Huang Z (2018) The singularity in the state-based peridynamic solution of uniaxial tension. Theor Appl Mech Lett \textbf{8}, 351--354. \doi{10.1016/j.taml.2018.05.008}

\bibitem[{Kim et~al.(2023)Kim, Bhalla, and Griffith}]{kim2023}
Kim KH, Bhalla APS, Griffith BE (2023) An immersed peridynamics model of fluid-structure interaction accounting for material damage and failure. J Comput Phys \textbf{493}, 112466. \doi{10.1016/j.jcp.2023.112466}

\bibitem[{Kraus(1967)}]{kraus1967}
Kraus H (1967) Thin Elastic Shells. John Wiley {\&} Sons, New York, NY

\bibitem[{Mikata(2012)}]{mikata2012}
Mikata Y (2012) Analytical solutions of peristatic and peridynamic problems for a {1D} infinite rod. Int J Solids Struct \textbf{49}, 2887--2897. \doi{10.1016/j.ijsolstr.2012.02.012}

\bibitem[{Mikata(2023)}]{mikata2023}
Mikata Y (2023) Analytical solutions of peristatics and peridynamics for {3D} isotropic materials. Eur J Mech A/Solids \textbf{101}, 104978. \doi{10.1016/j.euromechsol.2023.104978}

\bibitem[{Naumenko and Eremeyev(2022)}]{naumenko2022}
Naumenko K, Eremeyev VA (2022) A non-linear direct peridynamics plate theory. Compos Struct \textbf{279}, 114728. \doi{10.1016/j.compstruct.2021.114728}

\bibitem[{Nishawala and Ostoja-Starzewski(2017)}]{nishawala2017}
Nishawala VV, Ostoja-Starzewski M (2017) Peristatic solutions for finite one- and two-dimensional systems. Math Mech Solids \textbf{22}, 1639--1653. \doi{10.1177/1081286516641180}

\bibitem[{{O'Grady} and Foster(2014{\natexlab{a}})}]{ogrady2014a}
{O'Grady} J, Foster J (2014{\natexlab{a}}) Peridynamic beams: A non-ordinary, state-based model. Int J Solids Struct \textbf{51}, 3177--3183. \doi{10.1016/j.ijsolstr.2014.05.014}

\bibitem[{{O'Grady} and Foster(2014{\natexlab{b}})}]{ogrady2014b}
{O'Grady} J, Foster J (2014{\natexlab{b}}) Peridynamic plates and flat shells: A non-ordinary, state-based model. Int J Solids Struct \textbf{51}, 4572--4579. \doi{10.1016/j.ijsolstr.2014.09.003}

\bibitem[{Papanicolaou(2003)}]{papanicolaou2003}
Papanicolaou NC (2003) {A Galerkin Spectral Method for Fourth-Order Boundary Value Problems}. PhD thesis, University of Louisiana at Lafayette. \urlprefix\url{https://www.proquest.com/dissertations-theses/galerkin-spectral-method-fourth-order-boundary/docview/305212891/se-2}

\bibitem[{Papanicolaou and Christov(2023)}]{NectarIvan2023}
Papanicolaou NC, Christov IC (2023) Orthonormal eigenfunction expansions for sixth-order boundary value problems. J Phys: Conf Ser \textbf{2675}, 012016. \doi{10.1088/1742-6596/2675/1/012016}, \eprint{2308.00673}

\bibitem[{Papanicolaou et~al.(2009)Papanicolaou, Christov, and Homsy}]{papanicolaou2009}
Papanicolaou NC, Christov CI, Homsy GM (2009) Galerkin technique based on beam functions in application to the parametric instability of thermal convection in a vertical slot. Int J Num Meth Fluids \textbf{59}, 945--965. \doi{10.1002/fld.1845}

\bibitem[{Porubov et~al.(2018)Porubov, Osokina, and Michelitsch}]{porubov2018}
Porubov AV, Osokina AE, Michelitsch TM (2018) Nonlocal approach to square lattice dynamics. In: Altenbach H, Pouget J, Rousseau M, Collet B, Michelitsch T (eds) Generalized Models and Non-classical Approaches in Complex Materials 1, Springer International Publishing, Cham, Switzerland, pp 641--654. \doi{10.1007/978-3-319-72440-9\_34}

\bibitem[{Silling(2000)}]{silling2000}
Silling SA (2000) Reformulation of elasticity theory for discontinuities and long-range forces. J Mech Phys Solids \textbf{48}, 175--209. \doi{10.1016/S0022-5096(99)00029-0}

\bibitem[{Silling and Bobaru(2005)}]{silling2005}
Silling SA, Bobaru F (2005) Peridynamic modeling of membranes and fibers. Int J Non-Linear Mech \textbf{40}, 395--409. \doi{10.1016/j.ijnonlinmec.2004.08.004}

\bibitem[{Silling and Lehoucq(2010)}]{silling2010}
Silling SA, Lehoucq RB (2010) Peridynamic theory of solid mechanics. Adv Appl Mech \textbf{44}, 73--168. \doi{10.1016/S0065-2156(10)44002-8}

\bibitem[{Strutt(1877)}]{strutt1877}
Strutt JW (1877) The Theory of Sound, vol~1. Macmillan and Co., London. \urlprefix\url{http://books.google.com/books?id=GyI5AAAAMAAJ&oe=UTF-8}

\bibitem[{Taylor and Steigmann(2015)}]{taylor2015}
Taylor M, Steigmann DJ (2015) A two-dimensional peridynamic model for thin plates. Math Mech Solids \textbf{20}, 998--1010. \doi{10.1177/1081286513512925}

\bibitem[{Timoshenko and Woinowsky-Krieger(1959)}]{timoshenko1959}
Timoshenko S, Woinowsky-Krieger S (1959) Theory of Plates and Shells, 2nd edn. McGraw-Hill, New York, NY

\bibitem[{Yang et~al.(2020)Yang, Oterkus, and Oterkus}]{yang2020}
Yang Z, Oterkus E, Oterkus S (2020) A state-based peridynamic formulation for functionally graded {E}uler-{B}ernoulli beams. Comput Model Eng Sci \textbf{124}, 527--544. \doi{10.32604/cmes.2020.010804}

\bibitem[{Yang et~al.(2022{\natexlab{a}})Yang, Naumenko, Altenbach, Ma, Oterkus, and Oterkus}]{yang2022a}
Yang Z, Naumenko K, Altenbach H, Ma CC, Oterkus E, Oterkus S (2022{\natexlab{a}}) Some analytical solutions to peridynamic beam equations. Z Angew Math Mech (ZAMM) \textbf{102}, e202200132. \doi{10.1002/zamm.202200132}

\bibitem[{Yang et~al.(2022{\natexlab{b}})Yang, Naumenko, Ma, Altenbach, Oterkus, and Oterkus}]{yang2022b}
Yang Z, Naumenko K, Ma CC, Altenbach H, Oterkus E, Oterkus S (2022{\natexlab{b}}) Some closed form series solutions to peridynamic plate equations. Mech Res Commun \textbf{126}, 104000. \doi{j.mechrescom.2022.104000}

\bibitem[{Yang et~al.(2023{\natexlab{a}})Yang, Naumenko, Ma, and Chen}]{yang2023a}
Yang Z, Naumenko K, Ma CC, Chen Y (2023{\natexlab{a}}) Closed-form analytical solutions for the deflection of elastic beams in a peridynamic framework. Appl Sci \textbf{13}, 10025. \doi{10.3390/app131810025}

\bibitem[{Yang et~al.(2023{\natexlab{b}})Yang, Naumenko, Ma, Oterkus, and Oterkus}]{yang2023b}
Yang Z, Naumenko K, Ma CC, Oterkus E, Oterkus S (2023{\natexlab{b}}) Peridynamic analysis of curved elastic beams. Eur J Mech A/Solids \textbf{101}, 105075. \doi{10.1016/j.euromechsol.2023.105075}

\end{thebibliography}
}

\end{document}